\documentclass[12pt,preprint]{aastex}  

\usepackage{epsfig}
\usepackage{natbib}

\shorttitle{Rossby Vortex Instability with Toroidal Magnetic Fields}

\shortauthors{}

\begin{document}

\title{Nonaxisymmetric Rossby Vortex Instability with Toroidal Magnetic Fields 
       in Radially Structured Disks}

\author{Cong Yu\altaffilmark{1,2} and Hui Li\altaffilmark{2}}
\altaffiltext{1}{National Astronomical Observatories/Yunnan Astronomical Observatory, 
Chinese Academy of Sciences, Kunming, 650011}
\altaffiltext{2}{Los Alamos National Laboratory, Los Alamos, NM 87545;
 {\tt congyu@lanl.gov; hli@lanl.gov}}
\date{$\qquad\qquad\qquad\qquad\qquad \qquad\qquad$}  
\begin{abstract}
  We investigate the global nonaxisymmetric 
  Rossby vortex instability in a differentially rotating, 
  compressible magnetized accretion disk with radial density structures.
  Equilibrium magnetic fields are assumed to have only the toroidal component.
  Using linear theory analysis, we show 
  that the density structure can be unstable to nonaxisymmetric modes.
  We find that, for the magnetic field profiles we have studied, 
  magnetic fields always provide a stabilizing effect to the unstable 
  Rossby vortex instability 
  modes.
  We discuss the physical mechanism of 
  this stabilizing
  effect. 
  The threshold and properties of the unstable modes are also discussed in 
  detail.
  In addition, we present linear stability results for the global
  magnetorotational instability when the disk is compressible.


\end{abstract}



\section{INTRODUCTION}
Radial density structures can be common in astrophysical disks. 
For example, it is thought that a transition region exists in protostellar disks that can be 
caused by different degrees of coupling of the magnetic field with the disk
material (Gammie 1996; Terquem 2008). 
Tidal interactions between protoplanets and protoplanetary disks can
also give rise to dips/gaps in protoplanetary disks (Goldreich \& Tremaine 1980; Lin \& Papaloizou 1986; 
Ward 1997).
At the inner edge of black hole accretion disk, the sharp density contrast between the plunge region 
and the accretion disk can also be treated as a density transition structure.
 
 

Rossby wave instability or Rossby vortex instability (RVI) in thin Keplerian disks with density structures 
in the hydrodynamic limit has been studied in the linear theory
(Lovelace et al. 1999; Li et al. 2000). The existence of unstable modes has been found to
be associated with the radial gradients of an entropy-modified potential vorticity profile. 
RVI in disks will form vortices and shocks in the nonlinear limit (Li et al. 2001). 
Recently, RVI associated with a dip/gap is studied by de Val-Borro et al. (2007).
The intrinsic mechanism of the instability is the corotation amplification
caused by over-reflection (Goldreich et al. 1986). Unlike Papaloizou \& Pringle (1984) instability, 
the RVI does not depend on the reflection boundary conditions 
(see Li et al. 2000 for a detailed discussion). 

Magnetic fields are supposed to be present in accretion disks and would
greatly change the dynamics of disks. 
Previous magnetized disk instability analyses, both local and global, have mainly considered smooth disks 
without structures (e.g., Balbus \& Hawley 1991; Ogilvie \& Pringle 1996).
Global magnetorotational instability (MRI) with vertical and azimuthal magnetic fields have been studied by
Curry \& Pudritz (1996) and Ogilvie \& Pringle (1996), respectively.
They both found that global MRI are localized and Curry \& Pudritz (1996) gave an explanation in terms of
the Alfven resonance positions in the disk.
Curry \& Pudritz (1996) also found that global MRI is very sensitive to the boundary conditions.
Recently Pino \& Mahajan (2009) propose to reduce the dependence on boundary condtions by restricting
the rotation rate change only in a narrow range in radius.
Magnetized disk instability with step-shaped density structure has been considered by Tagger \& Pellat (1999), but 
they just considered purely poloidal magnetic fields. 
This kind of accretion-ejection instability only occurs in strongly magnetized disks with plasma $\beta\leq 1$.
The growth rates are typically small, $\sim$ several percent of the Keplerian frequency.
Due to the disk differential rotation,
the dominant component of magnetic fields should be toroidal (Balbus \& Hawley 1998).
How would the toroidal magnetic fields change the behavior of the RVI 
is still an open question. 

In this paper, we make such an attempt to study the magnetic field effects on structured disks
and extend the RVI study into disks with toroidal magnetic fields.
The paper is organized as follows. In \S 2 we describe the equilibrium disk with density
structures. In \S 3, we give the linear analysis of the magnetized RVI. 
We also discuss the effects of compressibility on global MRI.
Conclusions are given in \S 4.

\section{Disk Equilibrium and Structure Profile}
We use a cylindrical $(r,\phi,z)$ coordinate system. 
The equilibrium disk
is axisymmetric and in steady state, with unperturbed velocity ${\bf v}_0 = v_{\phi}{\hat{\bf e}_{\phi}}$.
The vertical stratification is neglected and is assumed to be uniform.
Only the azimuthal component of the magnetic fields is present. 
For the axisymmetric equilibrium disk, the radial force balance reads
\begin{equation}
\frac{v_{\phi}^2}{r} \equiv r\Omega^2 = 
\frac{1}{\rho}\left[\frac{d (p + \frac{1}{2} B_{\phi}^2) }{d r} + \frac{B_{\phi}^2}{r}\right] + \frac{d \Phi}{d r} \ ,
\end{equation}
where $\Phi$ is the gravitational potential of the central object and the 
disk self-gravity is not considered in the present paper. Here $\rho$, $p$ and $B_{\phi}$ 
are mass density, gas pressure and toroidal magnetic field, respectively.
We define $\Omega_0$ to be the Keplerian
angular velocity at $r_0$ and $v_0 = r_0 \Omega_0$. The length, time and mass density scale are normalized 
by $r_0$, $\Omega_0^{-1}$ and $\rho_0$, respectively. The disk is assumed to be isothermal with
a constant temperature. 
The isothermal sound speed is chosen as $ c_s $, and $p_0 = \rho_0 c_s^2$. 
The toroidal magnetic field is taken as
\begin{equation}
B_{\phi} = \lambda B_0 r^{-\alpha} \ ,
\end{equation}
where $B_0 = \sqrt{p_0}$. 
The strength and gradient of the toroidal magnetic field
is controlled by $\lambda$ and $\alpha$, respectively. We typically take $\alpha = 1$ in our paper.
At the characteristic radius $r_0$, the plasma $\beta$ in our formulation is approximately 
$\beta \sim p_0/(\lambda^2 B_0^2/2) = 2/\lambda^2$, but, in general, $\beta$ is a function of radius.

We focus in this paper on the configurations of disks with density dip/gap with 
a toroidal magnetic field and study the stability of such equilibrium disks.
For simplicity, we 
model the density dip/gap with a Gaussian profile
\begin{equation}
\rho = \rho_0\left\{1 - (G-1)\exp\left[-\frac{1}{2}\left(\frac{r-r_0}{\Delta }\right)^2\right]\right\} \ ,
\end{equation}
where $G$ ($1\leq G<2$) and $\Delta $ specify the amplitude and width of the density structure, respectively. 
The above profile is motivated by the fact that the protoplanet's tidal interaction with the gaseous disk
would induce a gap in the disk. 
The perturbation by a protoplanet leads to the excitation of spiral density waves at Lindblad resonances, which
carry an angular momentum flux. The waves deposit the angular momentum flux when the waves are dissipated. 
As a result the outer disk receives angular mometum from the protoplanet and the inner disk loses angular momentum
to the protoplanet. The outer (inner) disk gas gains (loses) angular momentum and moves outward (inward). 
When the tidal torque is greater than the viscous torque of the disk, a surface density dip/gap would be formed in
the vicinity of the protoplanet.
In numerical simulations the gap profile is similar to a Gaussian profile. 
Such configurations could also be relevant to the radial border of active zone and dead zone, 
where strong coupling of the magnetic field and disk material
gives rise to bigger accretion rate in the active zone compared to the dead zone. As a result, there would be
a density jump at the boundary between the active zone and the dead zone. The instability behavior of the
density jump at this border 
is quite similar to the outer edge of the 
density gap. For simplicity, we just consider the gap situation in this paper. 
Note that the instability behavior of the outer edge of the density dip/gap
is very similiar to the density jump case considered in Li et al. (2000).
We take $G=1.5$, $r_0 = 1$, $\Delta /r_0 = 0.05$ and $c_s = 0.07$.
The inner and outer radii of the disk are taken as $r_{\rm{in}} = 0.4$ and
$r_{\rm{out}} = 1.6$, respectively.
In Figure 1, we give the equilibrium disk with a dip/gap. The four panels 
are $P/P_0(r_0)$, $\Omega(r)/\Omega_K(r)$, $\kappa^2(r)/\Omega_K^2(r)$ and $\kappa^2/\Sigma\Omega$, 
respectively, where $\kappa$ is the radial epicyclic frequency and 
$\kappa^2=\frac{1}{r^3}\frac{d(\Omega^2 r^4)}{dr}$. The quantity $\kappa^2/(\Sigma\Omega)$ is the profile of 
potential vorticity (PV). We can observe that there are two minima at $r\simeq0.9$ and $1.1$ in the PV profile
related to the inner and outer edges of the gap, 
which imply that there may exist two unstable regions associated with both the inner
and outer edges.

\begin{figure}
\begin{center}
\epsfig{file=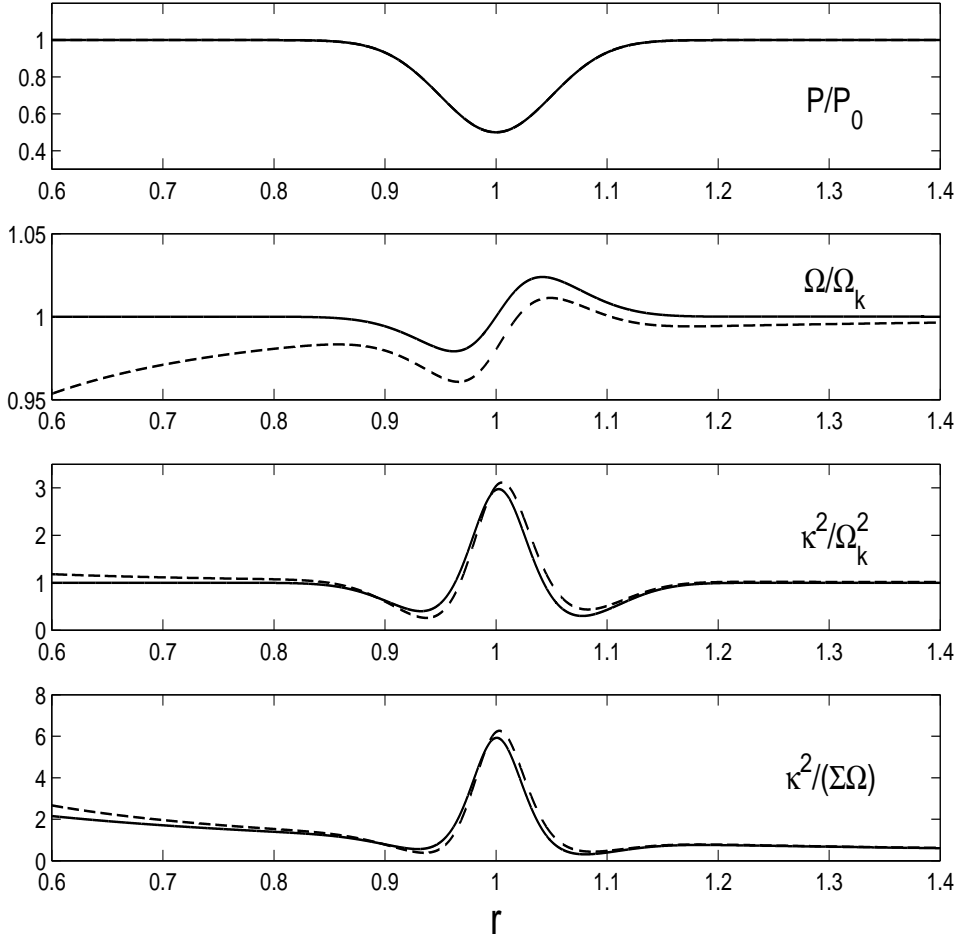,height=5in,width=5in,angle=0}
\end{center}
\caption{The equilibrium structure of the disk with a dip/gap.
        In this example the solid lines correspond to $\lambda = 0$ and $\alpha = 0$,
        the dashed line correspond to $\lambda = 2.0$ and $\alpha = 2$. Usually, we choose
        $\alpha = 1$, but when $\alpha = 1$, the dashed lines will overlap with solid lines.
        So we choose $\alpha = 2$ to make the differences appreciable.
        From top to bottom, the panels are 
        gas pressure, $P/P_0$, angular velocity, $\Omega/\Omega_K$, square of epicyclic frequency
        $\kappa^2/\Omega_K^2$, potential vorticity $\kappa^2/(\Sigma
        \Omega)$. In the first panel, the solid line and dashed line overlap.
        }
\end{figure}

\section{Linear Analysis of Vortex Instability}
Small perturbations to the inviscid compressible Euler equation are considered. The 
mass density is $\rho = \rho_0 + \delta \rho$, the 
gas pressure is $p = p_0 + \delta p$, the velocity is 
${\bf v} = {\bf v_0} + {\bf }u = \Omega r \hat{\bf e}_{\phi} + {\bf u}$ 
and ${\bf B} = B_{\phi}\hat{\bf e}_{\phi} + \delta{\bf B}$, 
where $\delta \rho$, $\delta p$, ${\bf u}$ and $\delta{\bf B}$ are perturbed mass density, 
gas pressure,  
velocity and magnetic field, respectively. 
The magnetohydrodynamic (MHD) equations of the  compressible disk are
\begin{equation}
\frac{D \rho}{D t} + \rho \nabla\cdot{\bf v} = 0 \ ,
\end{equation}
\begin{equation}
\frac{D {\bf v}}{D t} = - \frac{1}{\rho} \nabla p - \nabla \Phi + 
\frac{1}{\rho} \left(\nabla\times{\bf B}\right)  \times {\bf B}\ ,
\end{equation}
\begin{equation}
\frac{\partial {\bf B}}{\partial t} = 
\nabla\times \left({\bf v}\times{\bf B}\right)  \ .
\end{equation}
Note that the factor $4\pi$ is absorbed in the definition of magnetic field.

We linearize the above equations by taking perturbations proportional to 
$\exp(i m \phi + i k_z z - i \omega t)$, where $m = \pm 1, \pm 2, ...$ and $k_z$ are the azimuthal
and axial wavenumber, respectively, and $\omega = \omega_r + i \gamma$ is the mode eigenfrequency. 
We use
$\Psi = (\delta p + B_{\phi}\delta B_{\phi})/\rho$ as 
the perturbed total pressure divided by density 
($\delta B_{\phi}$ is the perturbed azimuthal magnetic field)  
and $\xi_r$ as the Lagrangian displacement. 
The linearized perturbation equations for $\xi_r$ and $\Psi$ read
\begin{equation}\label{linearper1} 
\frac{d \xi_r}{d r} = A_{11} \xi_r + A_{12} \Psi \ ,
\end{equation}
\begin{equation}\label{linearper2}
\frac{d \Psi}{d r} = A_{21} \xi_r + A_{22} \Psi \ .
\end{equation}
The four coefficients in the above two equations are 
\begin{equation}\label{equA11}
A_{11} =  - \left[ \frac{1}{L_1} + \frac{ k_{\phi}  c_s^2}{c_s^2 + c_a^2} 
\frac{( 2 \sigma \Omega + C_3 ) }{\sigma^2 -\sigma_M^2 } \right] \ ,
\end{equation}
%
\begin{equation}\label{equA12}
A_{12} = - \left[ \frac{1}{c_s^2 + c_a^2} 
- \frac{k_z^2}{\sigma^2 - \Omega_{a,z}^2} 
- \frac{ k_{\phi}^2}{\sigma^2 -\sigma_M^2} \left(\frac{c_s^2}{c_a^2+c_s^2}\right)^2 \right] \ ,
\end{equation}
%
\begin{equation}\label{equA21}
A_{21} = 
\sigma^2 - \kappa^2 - \Omega_{a,r}^2 
- \frac{4\Omega^2\sigma_M^2 + C_3 \Omega_{a,\phi}^2}{\sigma^2 - \sigma_M^2}
- \frac{2 (\Omega_{a,\phi}^2 + C_3) \sigma \Omega }{\sigma^2 -\sigma_M^2} \ ,
\end{equation}
%
and
\begin{equation}\label{equA22}
A_{22} =  
\left[ (2 \sigma \Omega + \Omega_{a,\phi}^2)
\frac{ k_{\phi}  \frac{c_s^2}{c_a^2+c_s^2} }{\sigma^2 -\sigma_M^2 } + \frac{1}{L_2}\right] \ ,
\end{equation}
where
\[
k_{\phi} = \frac{m}{r} \ , \quad c_a^2 = \frac{B_{\phi}^2}{\rho} \ , \quad \sigma = \omega - m\Omega \ ,
\quad \sigma_M^2 = \frac{k_{\phi}^2 c_s^2 c_a^2}{c_s^2 + c_a^2} \ ,
\]
\[
C_3 = k_{\phi} \left[ \frac{c_s^2}{c_a^2+c_s^2} C_1
+ \frac{ B_{\phi} }{\rho} \frac{d B_{\phi} }{d r} + \frac{c_a^2}{r} \right] \ ,
\]
and
\[
C_1 = c_a^2 \frac{d\ln\rho}{d r} + \frac{c_a^2}{r} - \frac{B_{\phi}}{\rho} \frac{d B_{\phi}}{d r} 
= - \frac{\rho r^2}{2}\frac{d}{dr} \left(\frac{B_{\phi}^2}{\rho^2 r^2}  \right) \ .
\]
The two length scales in the above equations (\ref{equA11}) and (\ref{equA22}) are defined by
\begin{equation}
1/L_1 \equiv 
\frac{c_s^2}{(c_s^2 + c_a^2)} \frac{1}{r} + \frac{c_s^2}{(c_s^2 + c_a^2)} \frac{d\ln\rho}{d r} 
+ \frac{1}{(c_s^2 + c_a^2)} \frac{B_{\phi}}{\rho} \frac{d B_{\phi}}{d r} \ ,
\end{equation}
and
\begin{equation}
1/L_{2} \equiv 
- \frac{d\ln\rho}{d r} +  C_2 - \frac{2 c_a^2}{c_a^2+c_s^2} \frac{1}{r} \ ,
\end{equation}
where
\[
C_2 = \frac{1}{\rho (c_s^2 + c_a^2)}\left(\frac{d (p + \frac{1}{2}B_{\phi}^2)}{d r} + \frac{B_{\phi}^2}{r} \right) 
= \frac{1}{(c_a^2 + c_s^2)}(\Omega^2 r - \nabla \Phi)\ .
\]
The three quanities $\Omega_{a,r}^2$, $\Omega_{a,\phi}^2$ and $\Omega_{a,z}^2$ 
in equations (\ref{equA12}) and (\ref{equA21}) are as follows,
\[
\Omega_{a,r}^2 \equiv 
k_{\phi}^2 c_a^2 +  C_1 C_2 + \frac{2 c_s^2}{c_a^2+c_s^2} \frac{C_1}{r} \ ,
\]
\[
\Omega_{a,\phi}^2 \equiv 
m\left( \frac{c_a^2 C_2}{r} + \frac{2 c_s^2 c_a^2}{(c_a^2+c_s^2) r^2}\right) \ ,
\]
\[
\Omega_{a,z}^2 \equiv k_{\phi}^2 c_a^2 \ .
\]
Note that in the axisymmetric incompressible disks with uniform density background, 
the quantity $\Omega_{a,r}^2$ reduces to $-\frac{r}{\rho}\frac{d}{d r}(B_{\phi}/r)^2$, 
which is the same as given by Chandrasekhar (1961). When taking the incompressible limit,
equations (\ref{linearper1}) and (\ref{linearper2}) reduce to (2.13) and (2.14) in Ogilvie \& Pringle (1996).
The two equations (\ref{linearper1})  and (\ref{linearper2}) can be combined to get a single 
second order differential equation with respect to $\Psi$, 
\begin{equation}\label{2ndorderode}
\Psi^{''} + B(r)\Psi^{'} + C(r) \Psi = 0 \ ,
\end{equation}
where
\[
B(r) = - \left(A_{11} + A_{22} - \frac{ A_{21}^{'}}{A_{21}}  \right) \ ,
\] 
\[
C(r) = - \left(A_{12}A_{21} - A_{11}A_{22} + A_{22}^{'} - \frac{A_{22} A_{21}^{'}}{A_{21}}  \right) \ ,
\]
and the prime denotes the derivative with respect to $r$. 

\subsection{Axisymmetric Stability}
Before we proceed to investigate the vortex instability produced by inflexion points in the PV
profile, we need to ensure the equilibrium is stable to axisymmetric perturbations.
To this end, we use the sufficient condition for local stability, 
i.e., the generalized Rayleigh criterion
\begin{equation}\label{shcriterion}
\kappa^2(r) + \Omega_{a,r}^2 \geqslant 0 \ . 
\end{equation}
This criterion can be readily seen in the incompressible axisymmetric uniform
density background model. In such a case the two coupled equations become (we temporarily
keep $k_z$ in this subsection in order to see the generalized Rayleigh criterion and suppress $k_z$ 
in the later RVI analysis)
\begin{equation}
D_{*}\xi_r =  \frac{k_z^2}{\sigma^2} \Psi \quad \hbox{and} \quad 
D \Psi = (\sigma^2-\kappa^2-\Omega_{a,r}^2) \xi_r \ ,
\end{equation}
where $D_*=D+\frac{1}{r}$ and $D=\frac{d}{dr}$. Eliminating $\Psi$ from 
the above two equations, we arrive at
\begin{equation}
(D D_{*} - k_z^2)\xi_r = - \frac{k_z^2}{\sigma^2}(\kappa^2 + \Omega_{a,r}^2) \xi_r
\end{equation}
Simple variational principle analysis gives that a necessary and sufficient condition for
axisymmetric perturbations to be stable is that $\kappa^2 + \Omega_{a,r}^2$ be positive
throughout the whole radius range (e.g., Chandrasekhar 1961).
In all of our following examples, we require that the above asxisymmetric stablility criterion is met.
We will see that the stronger the magnetic field, the bigger the quantity $\Omega_{a,r}^2$ and it is
this term that stabilizes the hydrodynamic RVI. 
%

\subsection{Method of Solving Linear Eigenvalue Problem}
Since the eigenfrequency $\omega$ is in general complex, equations 
(\ref{linearper1}) and (\ref{linearper2}) are a pair of
first order differential equation with complex coefficients which are functions 
of $r$.  
We choose to use the relaxation method to solve these equations (Press et al. 1992). In this method,
the ODEs are replaced by finite difference equations on a mesh of points covering the domain of interest.
The relaxation method needs an initial trial solution 
that can be improved by a Newton-Raphson technique.
Iterations are carried out by carefully designed Gaussian elimination adapted to block diagonal matrix.
After iterations the initial trial solution will gradually converge to the two point boundary eigenvalue problem. 
For the MRI calculation, we use rigid boundary conditions. 
For the RVI calculation, the boundary conditions implemented are such that waves propagate 
away from the density structure in both inner and outer parts of the disk (e.g., Li et al. 2000).

\subsection{3D Results of Linear Analysis on MRI}
We solved  equations (\ref{linearper1}) and (\ref{linearper2}) in two cases. 
One corresponds to a  three dimensional infinite cylinder with $k_z \ne 0$ and 
the other corresponds to two dimensional thin disk with $k_z = 0$.
In this section we focus on the MRI and we treat the disk
density as uniform (i.e. $G=1$) and $k_z \ne 0$. 
Only the global nonaxisymmetric MRI is considered.
The difference between our model and Ogilvie \& Pringle (1996) (hereafter OP)
is that they considered the incompressbile limit while our model is a compressible one.

\begin{figure}
\begin{center}
\epsfig{file=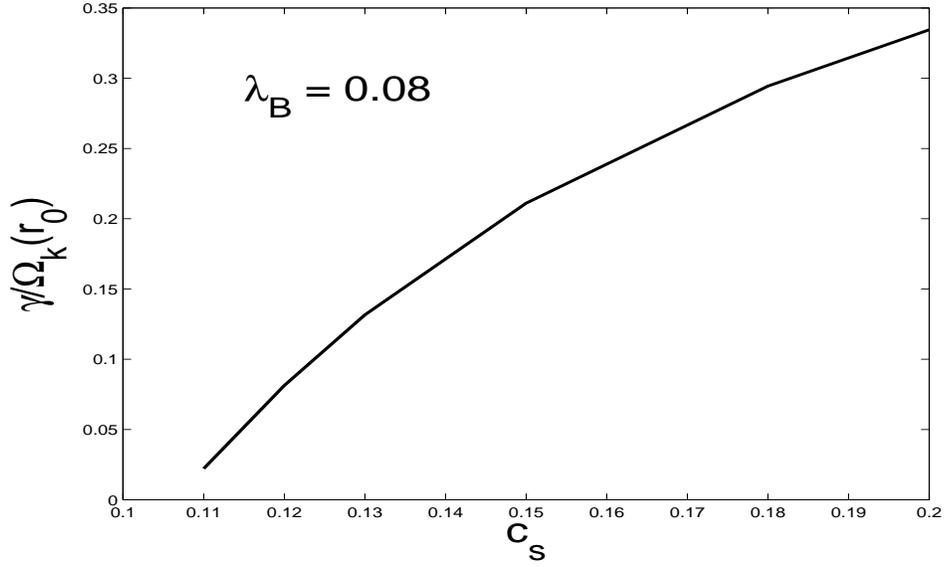,height=3in,width=5in,angle=0}
\end{center}
\caption{
         Dependence of growth rate of MRI on the compressbility for a fixed magnetic field $\lambda_B = 0.08$.
         $k_z = 28$ and $m = 5$ are chosen. 
        }
\label{csdependence}
\end{figure}

The equilibrium setup given by OP has, in normalized units,
$\rho_0 = 1$, $v_{\phi} = r^{1/2}$, $B_{\phi} = \lambda_B r^{-1}$, $\lambda_B$ is
a constant and taken as $0.2$.
Here $\lambda_B$ is related to $\lambda$ in equation (2) by 
$\lambda_B = \lambda\ c_s$. 
We take $k_z = 28$ and $m = 5$ in the following analysis 
for we can get relatively higher growth rate with this choice of parameters. 
In order to get to the realistic value of sound speed for disks, such as $c_s = 0.07$, we first recover 
OP's results by taking $c_s = 100$ and $\lambda_B = 0.2$. 
Then we gradually reduce the sound speed from $c_s = 100$ to $c_s = 0.5$ and keep $\lambda_B = 0.2$.
We find that when the sound speed reaches $c_s = 0.5$, we can not
find unstable modes any more. From a physical point of view, this means that the magnetic fields become 
too strong and the MRI is suppressed.
Then when we keep $c_s = 0.5$ and gradually reduce $\lambda_B$ from $0.2$ to  $0.1$, 
we can find unstable MRI again. 
After this, we keep $\lambda_B = 0.1 $ and gradually reduce the sound speed $c_s$ from $ 0.5$ to $ 0.2$. 
The MRI unstable modes growth rates decrease as we reduce the sound speed $c_s$ from $ 0.5$ to $ 0.2$.
Repeating the above process for several times, we gradually get to a realistic value of sound speed $c_s = 0.07$.
In Figure \ref{csdependence}, we show how the growth rate of MRI varies with the sound speed while
the magnetic field strengh is fixed with $\lambda_B = 0.08$.

\begin{figure}
\begin{center}
\epsfig{file=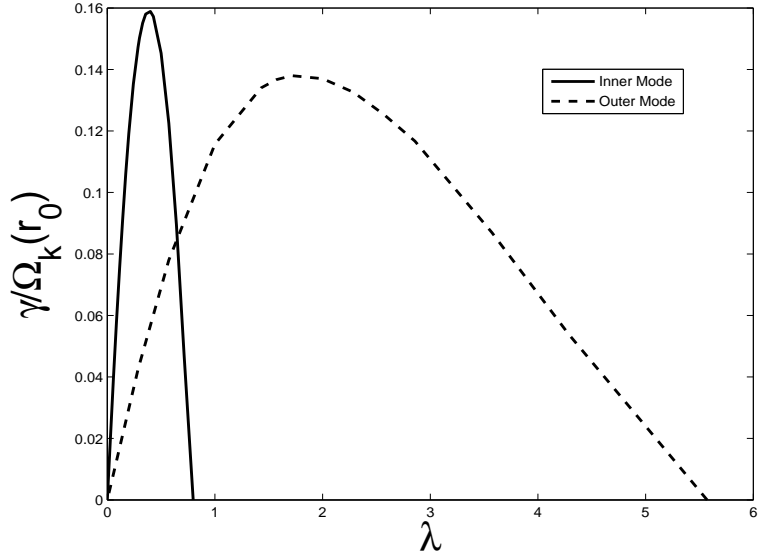,height=3in,width=4in,angle=0}
\end{center}
\caption{
         Dependence of the MRI growth rate $\gamma$ on the magnetic field 
         strength for the $m=5$ modes with $k_z=28.0$, $c_s = 0.07$. 
        }
\label{MRIlameffects}
\end{figure}

Once we find the unstable mode with the realistic sound speed for disks, we can slowly vary the magentic field 
strength to find the dependence of growth rates on magnetic field.
Figure \ref{MRIlameffects} gives the effects of the magnetic field strength on the growth rate 
for a particular mode with $k_z=28$ and $m=5$ for $c_s = 0.07$.
Most unstable modes are achieved with moderate magnetic field strength. 
This is different from the behavior of RVI, whose growth rate is a monotonically decreasing 
function of the magnetic field strength (see \S 4.3.2). 
And the peaks are achieved at different magnetic field strength for the inner and outer modes. 
The inner mode reaches the greatest growth rate around $\lambda=0.4$ or $\lambda_B = 0.028$ 
and the outer mode around $\lambda=1.71$ or $\lambda_B = 0.12$.
Surprisingly, we find that the plasma $\beta < 1$ for the high $\lambda$ situation
in Figure \ref{MRIlameffects}. 
Usually MRI is suppressed in low $\beta$ magnetized gas, here we can see that gas compressibility 
extends the MRI to the low $\beta$ regime.

\begin{figure}
\begin{center}
\epsfig{file=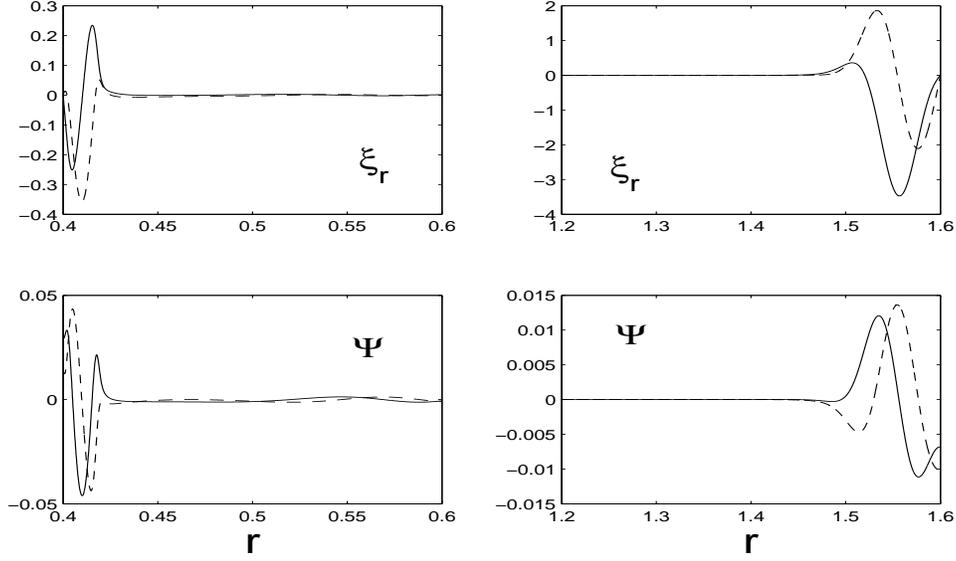,height=3in,width=5in,angle=0}
\end{center}
\caption{
         The left (right) two panels show the inner (outer) mode eigenfunction $\xi_r$ and $\Psi$. 
         The real part is shown by the solid line, the imaginary part by dashed line. 
         The inner mode is mainly between inner boundary $r_{\rm{in}} = 0.4$
         and $r_2=0.42$ while the outer mode is between $r_1=1.46$ and the 
         outer boundary $r_{\rm{out}} = 1.6$. 
        }
\label{localized}
\end{figure}

\begin{figure}
\begin{center}
\epsfig{file=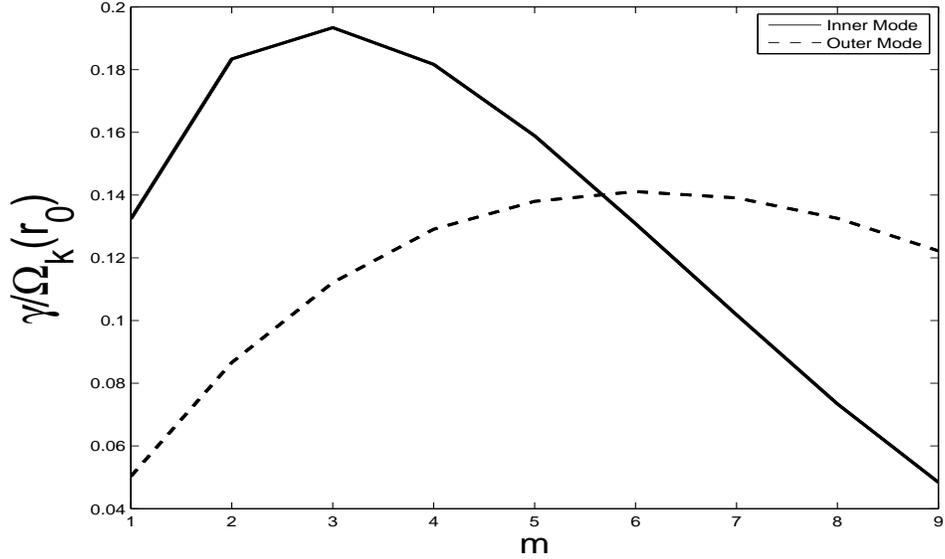,height=3in,width=5in,angle=0}
\end{center}
\caption{
         Dependence of the MRI growth rate on the azimuthal wave number $m$. Solid line is 
         for the inner mode, dashed line for the outer mode. 
        }
\label{MRImdependence}
\end{figure}

The eigenfunctions of unstable modes 
are quite similar to the results of Ogilvie \& Pringle (1996), the compressible global MRI modes are also
localized at the boundaries, which has been explained by Curry \& Pudritz (1996) in
terms of Alfven resonance positions in the disk.
Figure \ref{localized} shows examples of the MRI eigenfunction of the above mentioned two most ustable modes.
We find that the amplitudes concentrate either at the inner or at the outer boundary 
(that is why these modes are named).
The most unstable inner mode frequency is $\omega_r/(m\Omega_0) = 3.8077$, the radial postions where  
$\omega-m\Omega(r) = \pm k_{\phi} c_a(r)$ are $r_1 = 0.3979$ and $r_2 = 0.4219$ and the corotaion radius
is at $r_{c1} = 0.4102$. We note that $r_1$ is outside the computaional domain, and the amplitude is 
mainly confined between the inner boundary $r_{\rm{in}} = 0.4$ and $r_2$. 
The most unstable outer modes frequency is $\omega_r/(m\Omega_0)=0.5124$. The radial positions where
$\omega-m\Omega(r) = \pm k_{\phi} c_a(r)$ are $r_3 = 1.4564$ and $r_4 = 1.6573$ and 
the corotaion radius is at $r_{c2} = 1.5617$. 
We note that, in this case, $r_4$ is outside the computational domain.
The amplitude is mainly confined between $r_3$ and the outer boundary $r_{\rm{out}} = 1.6$.
These observations are essentially the same as the results of Curry \& Pudritz (1996).
We expect that the highly localized MRI would not affect the RVI much even though it has 
greater growth rates than RVI, because the radial density structure is far from the disk boundaries.
We have also investigated the azimuthal wave number dependence of the MRI growth rate.
Figure \ref{MRImdependence} gives the variation of growth rate with the azimuthal wave number $m$ 
of these two modes.
The inner $m=3$ mode and outer $m=6$ mode have the greatest growth rates.
We use the rigid boundary conditions $\xi_r = 0$ at both boundaries. 
We also tried the outflow boundary conditions used in the RVI calculation,
but we  can not find unstable modes any more. So global MRI is quite sensitive to boundary 
conditions as shown by other studies as well (e.g., Curry \& Pudritz 1996). 


\subsection{2D Results of Linear Analysis on Rossby Vortex Instability }
For disks with radial density structures, we restrict the analysis in the 2D $(r,\phi)$ plane $(k_z = 0)$.
In this limit, MRI is not excited. For the following analysis, we will focus on RVI with magnetic fields.
\subsubsection{Representative hydrodynamic examples}\label{represent}
\begin{figure}
\begin{center}
\epsfig{file=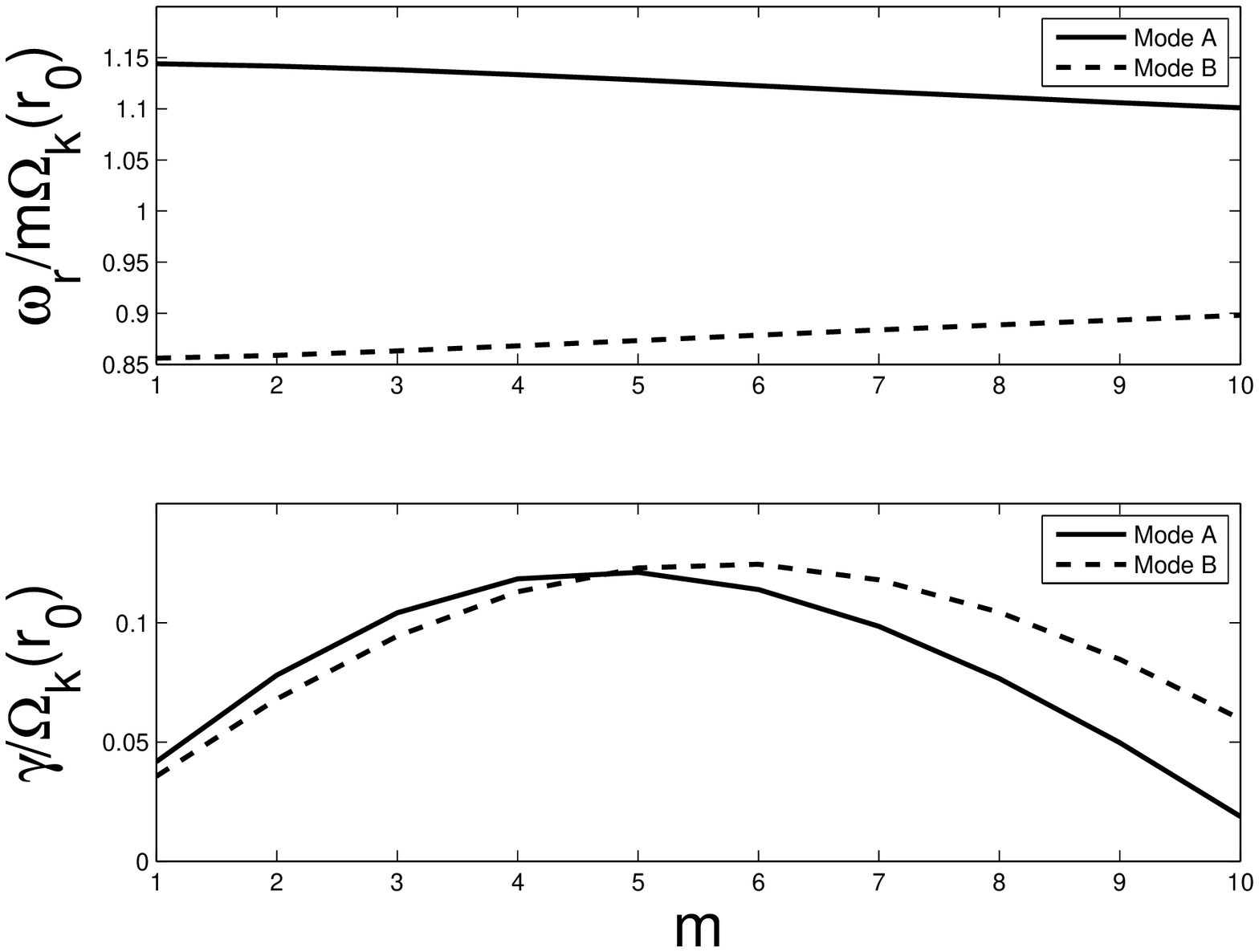,height=3in,width=5in,angle=0}
\end{center}
\caption{Dependence of the RVI mode frequency $\omega_r$ and  growth rate $\gamma$
         on the azimuthal mode number $m$.
         The solid and dashed line are for the inner edge (A) and outer edge mode
         (B), respectively. $B_{\phi}$ is zero in this case.
        }
\label{hydroeigenvalue}
\end{figure}

We first show the hydrodynamic results for a representative case with $G=1.5$,
$c_s = 0.07$ and $\Delta /r_0 = 0.05$. Solving equations (\ref{linearper1}) 
and (\ref{linearper2}), we find many unstable modes. 
Figure \ref{hydroeigenvalue} shows the dependence of 
mode frequency $\omega_r$ and growth rate $\gamma$
on the azimuthal mode number $m$ for the inner edge mode 
and the outer edge mode, respectively. 
The overall behaviour of the hydrodynamic instability is in good agreement with results 
of Li et al. (2000) and  de Val-Borro et al. (2007).

\subsubsection{Results of RVI with Magnetic Fields}
\begin{figure}
\begin{center}
\epsfig{file=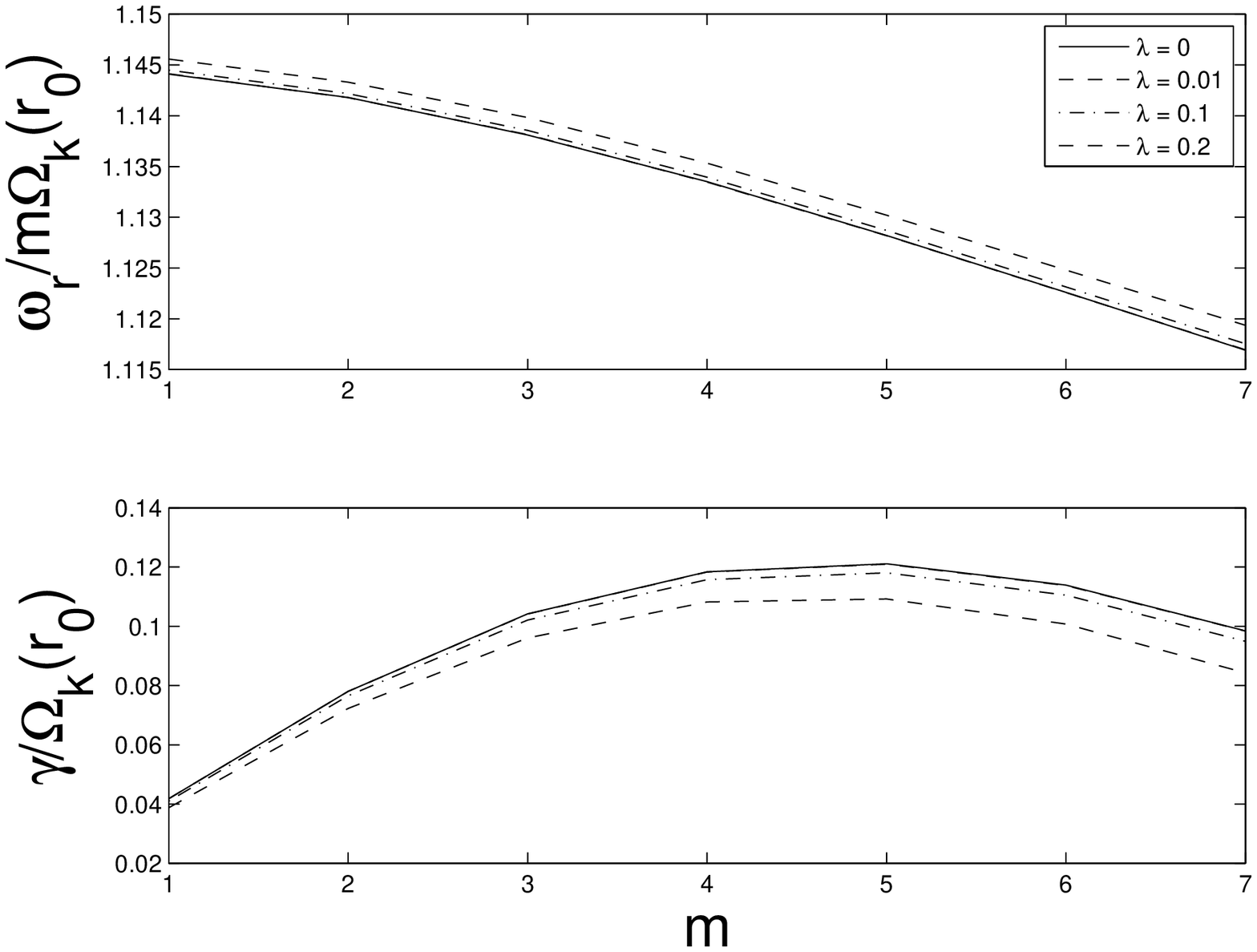,height=3in,width=5in,angle=0}
\end{center}
\caption{Dependence of the mode frequency $\omega_r$ and growth rate $\gamma$
         on the azimuthal mode number $m$ for the inner edge mode with different
         strength of magnetic field. $\alpha$ is set to be 1. The grow rate decreases with
         the strength of magnetic field, but the frequency increases slightly
         with the strength of the magnetic field. For small magnetic field
         $\lambda=0.01$, it is indistinguishable from the nonmagnetic case.
        }
\label{eigenvalueA}
\end{figure}

\begin{figure}
\begin{center}
\epsfig{file=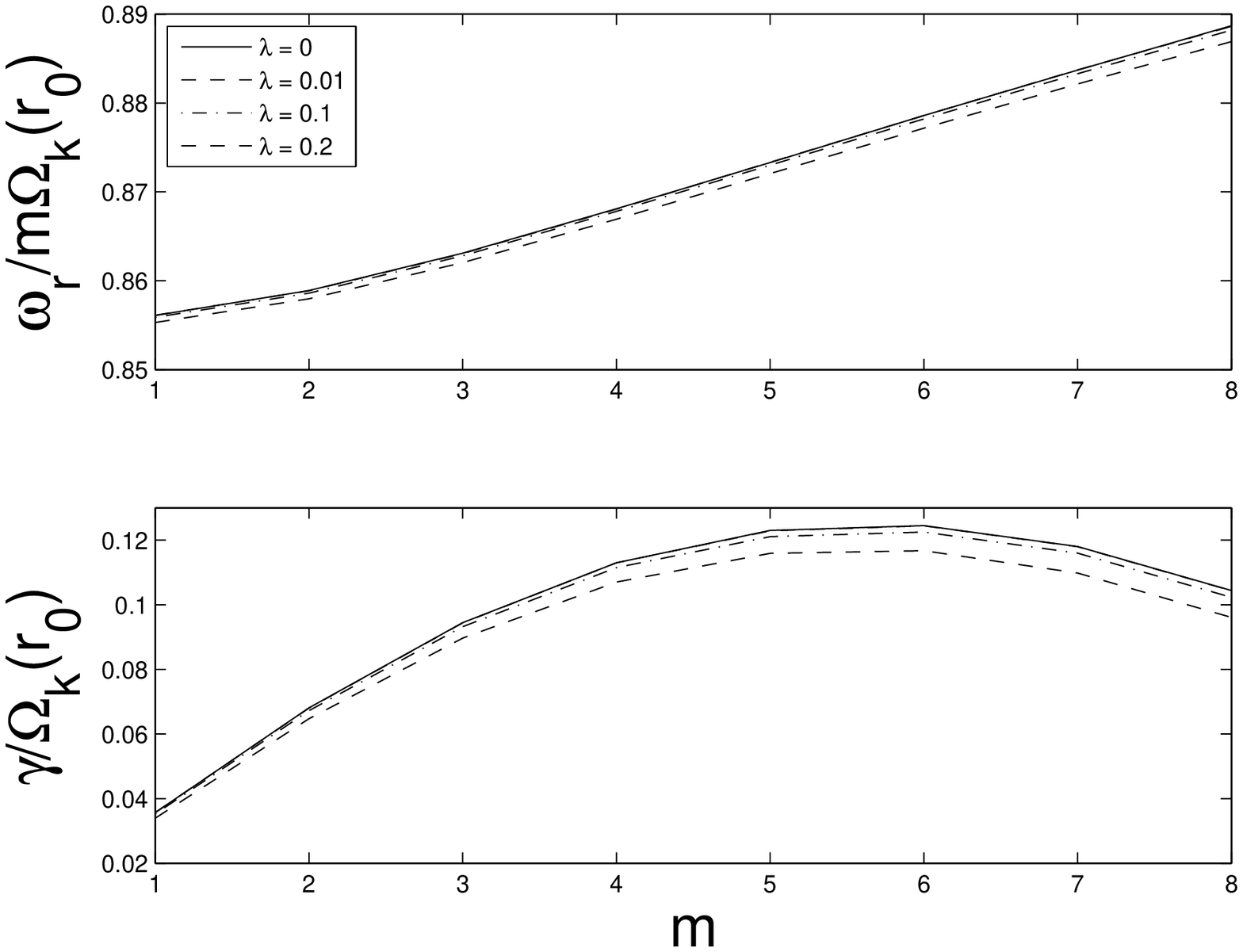,height=3in,width=5in,angle=0}
\end{center}
\caption{
          Same as Fig. \ref{eigenvalueA} but for the outer edge mode.
        }
\label{eigenvalueB}
\end{figure}

\begin{figure}
\begin{center}
\epsfig{file=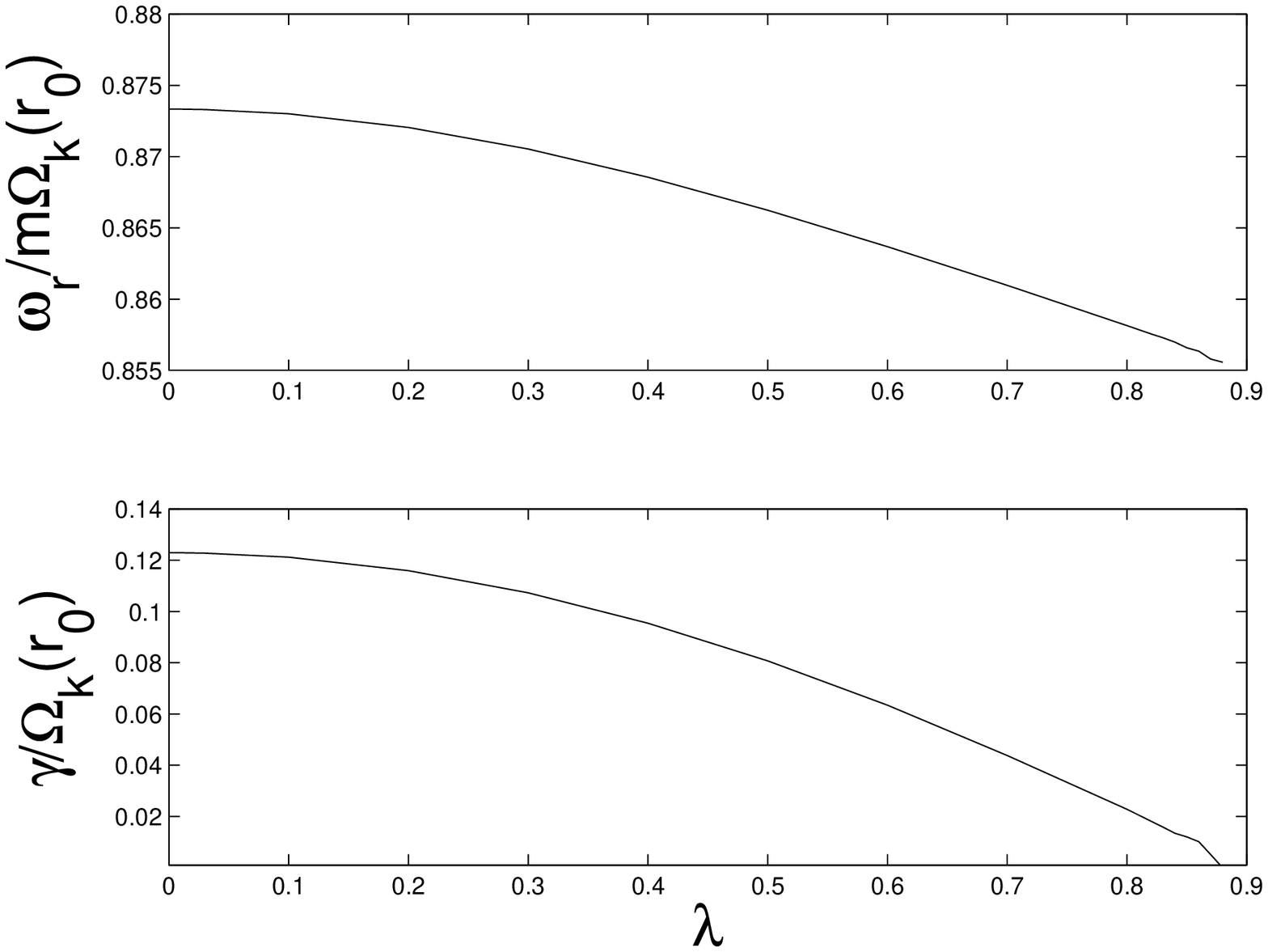,height=3in,width=5in,angle=0}
\end{center}
\caption{Dependence of the $m=5$ outer edge mode
         frequency $\omega_r$ and  growth rate $\gamma$
         on the strength of magnetic field. As the magnetic field increases,
         the growth rate diminishes. When $\lambda \sim 0.88$, or plasma 
         $\beta \sim 2/\lambda^2 \sim 2.6$, the RVI is 
         suppressed.
        }
\label{lamformeq5}
\end{figure}

With the inclusion of magnetic fields, we calculate the growth rate of the unstable 
modes.
In Figure \ref{eigenvalueA} and \ref{eigenvalueB}, we present the results
of the effects of different field strengths on the frequency and growth rate for different
azimuthal wavenumber modes associated with the inner and outer edge of the dip/gap, respectively.
The variation of the frequency and growth rate of the $m=5$ outer edge mode with
different magnetic field strength is presented in Figure \ref{lamformeq5}.
For this particular mode, both the frequency and growth rate decrease monotonically with the increase
of magnetic field. When the magnetic field is strong enough, we can see that the Rossby
vortex instability will be completely suppressed by the presence of magnetic field.
From Figure \ref{lamformeq5}, when the parameter $\lambda \sim 0.88$, or when the plasma $\beta$ is
approximately $2/\lambda^2 \sim 2.6$, the RVI is almost completely suppressed.

\begin{figure}
\begin{center}
\epsfig{file=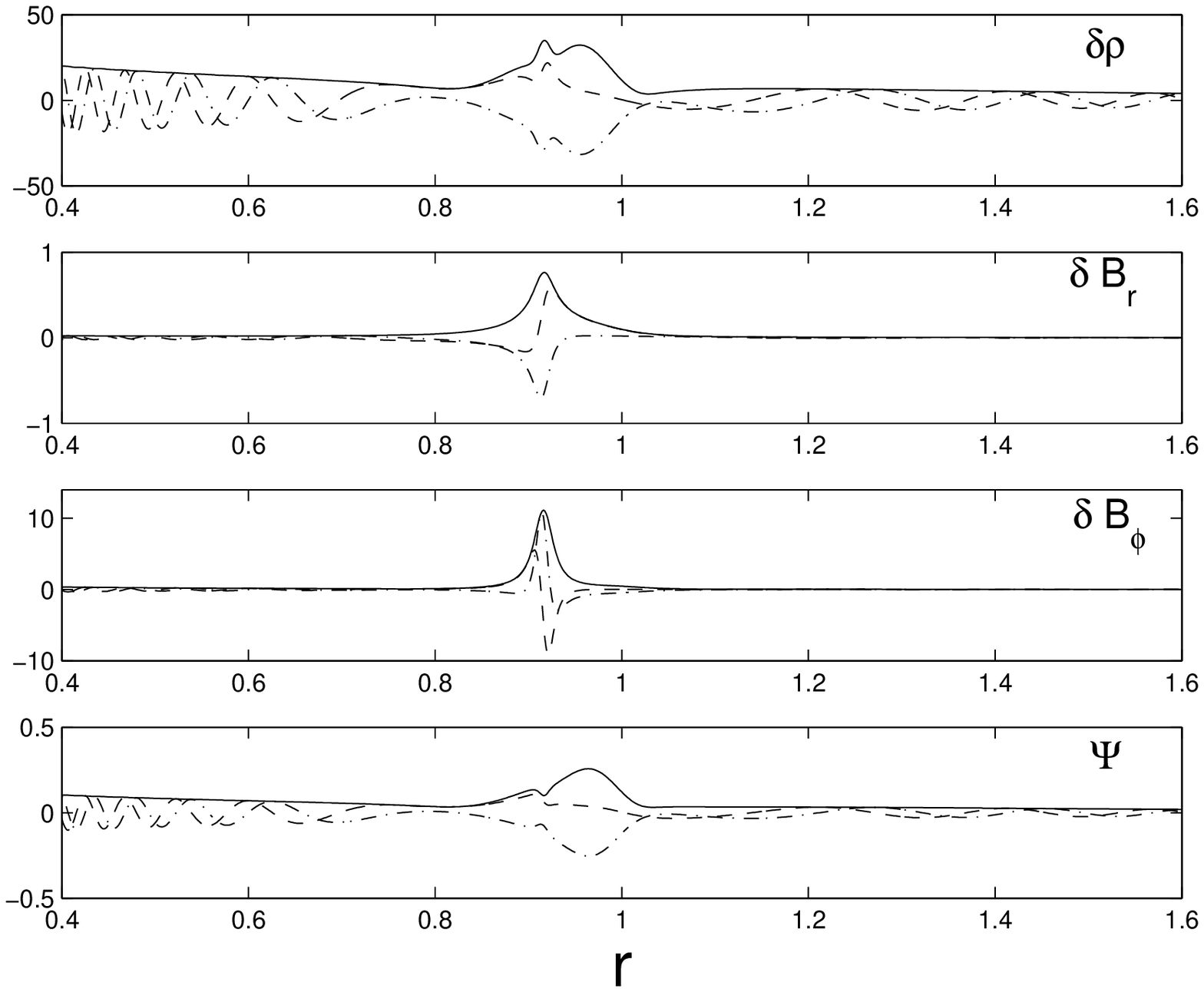,height=4in,width=4in,angle=0}
\end{center}
\caption{
         Eigenfunction for the inner edge mode (Mode A) of RVI with magnetic fields $\lambda = 0.1$. 
         Shown are the perturbed density, 
         the radial and azimuthal magnetic perturbations and the
         perturbation function $\Psi$ for 
         $m = 5$. The dashed line is the real part, the dot-dashed line is the
         imaginary part, and the solid line is the amplitude.
         }
\label{m5inner}
\end{figure}

\begin{figure}
\begin{center}
\epsfig{file=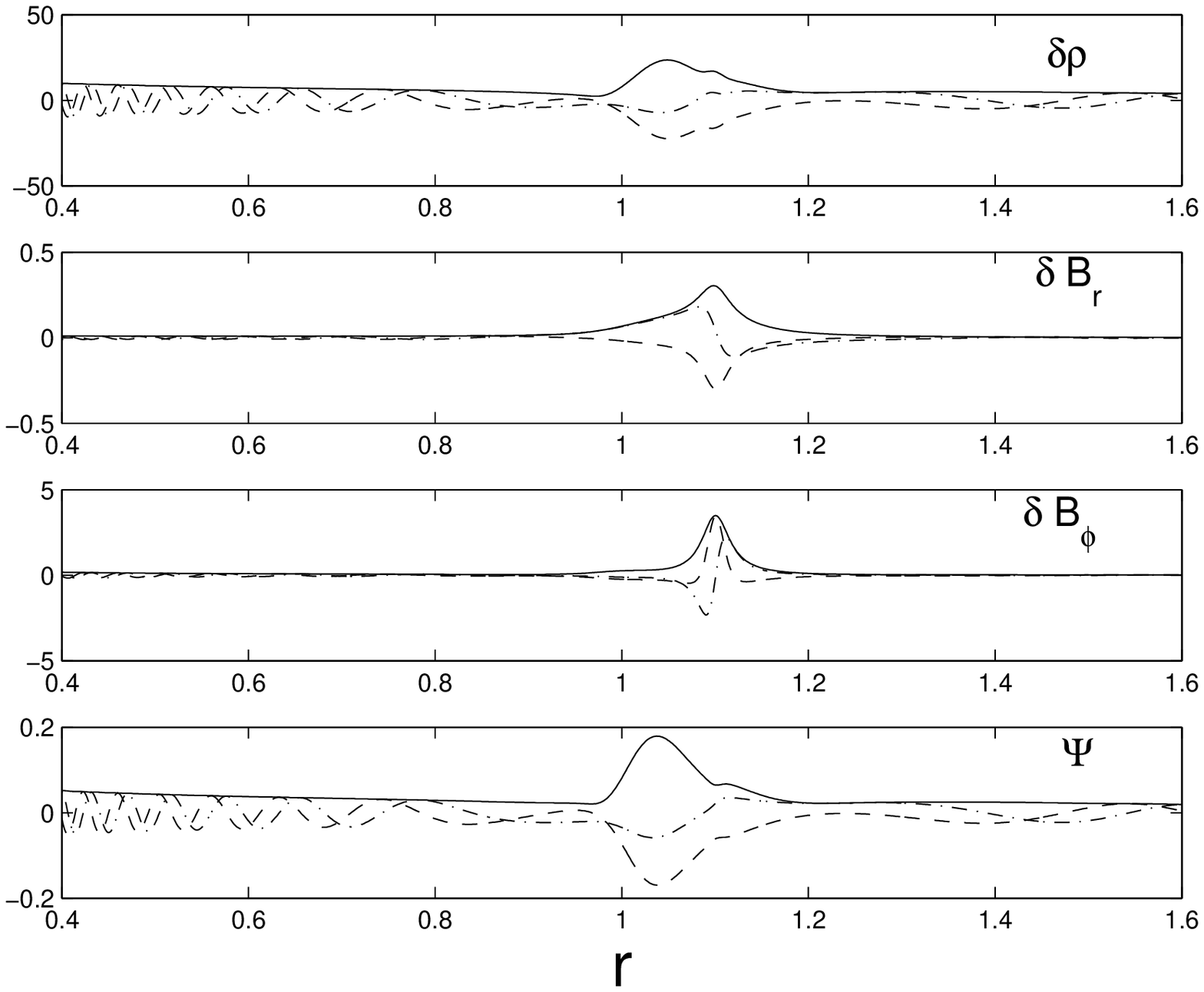,height=4in,width=4in,angle=0}
\end{center}
\caption{
          Same as Fig. \ref{m5inner} but for the outer edge mode.
         }
\label{m5outer}
\end{figure}

\begin{figure}
\begin{center}
\epsfig{file=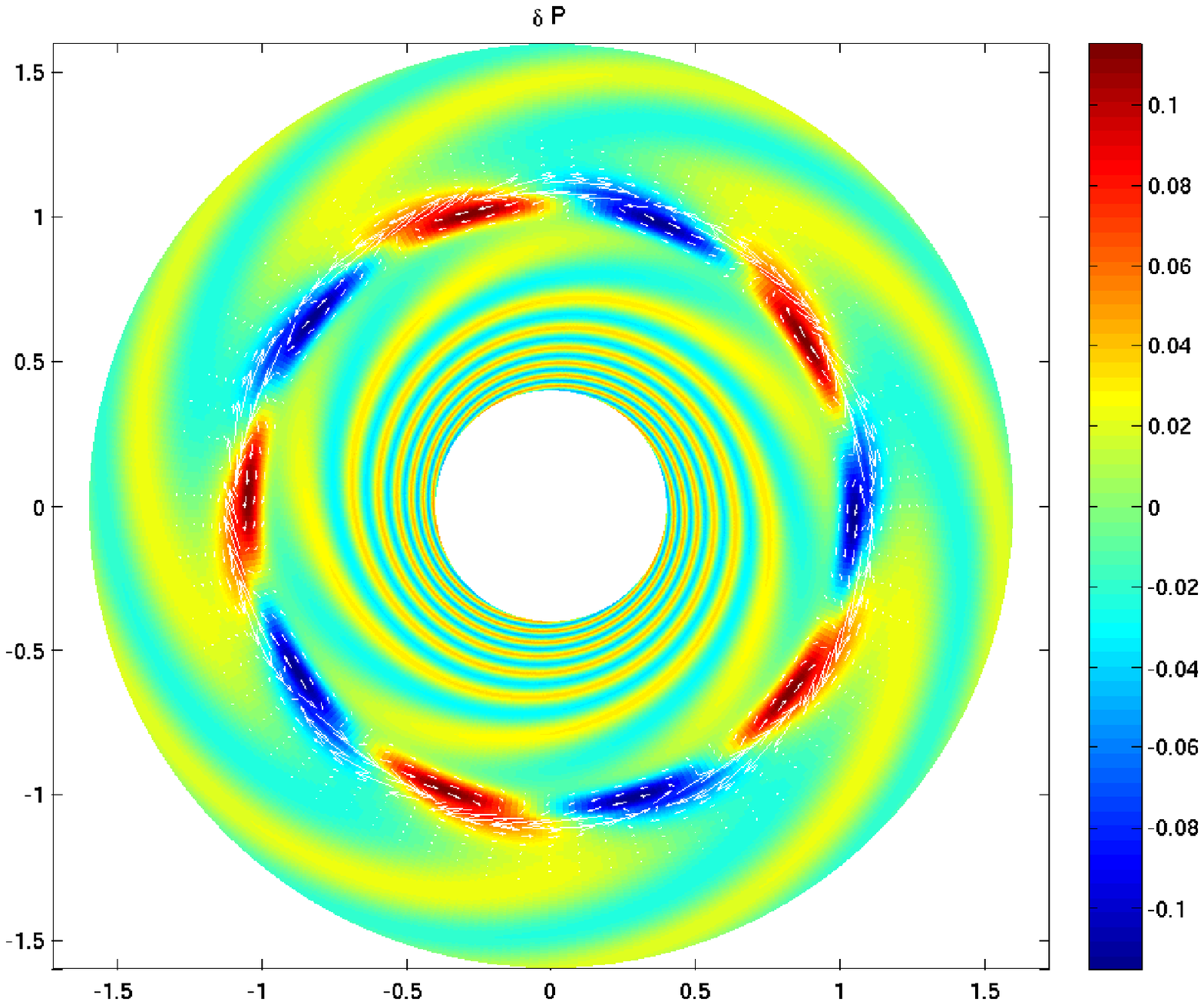,height=5in,width=6.2in,angle=0}
\end{center}
\caption{
         Two dimensional distribution of the gas pressure perturbation based on linear theory of  
         the outer edge $m = 5$ mode with $\lambda = 0.1$ and $\alpha = 1.0$.
         Arrows indicate the perturbation magnetic vector near $r_0$.
         Units are arbitrary.
        }
\label{deltap2d}
\end{figure}

The eigenfunctions of the $m=5$ unstable modes with $\lambda = 0.1$ are shown 
in Figure \ref{m5inner} and Figure \ref{m5outer} 
for the inner edge mode 
and outer edge mode, 
respectively. 
The unstable inner edge mode 
has a growth rate
$\gamma/\Omega_0 \approx 0.1181$ and a real frequency $\omega_r/(m\Omega_0)\approx 1.1287$.
The unstable outer edge mode 
has a growth rate 
$\gamma/\Omega_0 \approx 0.1211$ and a real frequency $\omega_r/(m\Omega_0)\approx 0.8730$.
We have used outward-propagating sound wave boundary conditions to obtain these eigenfunctions.
The relative phase shift between real and imaginary parts indicates this propagation. 
Two dimensional distribution of the perturbation are shown in 
Figure \ref{deltap2d}. 
We can identify that vortices 
develop around the radial density structure. 
The azimuthal pressure gradient is crucial for the formation of 
anticyclonic vortices (see Fig. $5$ in Li et al. (2001) for a detailed explanation).
When the equilibrium azimuthal magnetic fields are present, magnetic fields
restrict the development of radial velocity thus the formation of the vortices.
As a result the growth rate is reduced by the azimuthal magnetic fields.
 
\subsubsection{Stabilizing effect of Magnetic Fields and Physical Mechanism for the Instability}

As we mention before in the axisymmetric analysis, the quantity $\Omega_{a,r}^2$ plays an essential role
in stabilizing the RVI. 
Note that $\Omega_{a,r}^2$ can be written as:
\begin{equation}
\Omega_{a,r}^2 \equiv k_{\phi}^2 c_a^2 + c_a^2\left(\frac{d\ln\rho}{d r} +\frac{1+\alpha}{r}\right)
\left(\frac{c_s^2}{c_a^2+c_s^2}\frac{d\ln\rho}{d r} +\frac{(1-\alpha)c_a^2 + 2 c_s^2}{c_a^2+c_s^2} \frac{1}{r}\right) \ .
\end{equation}
Usually, the instability takes place near the density structure in the disk, where the density
scale length is much shorter than the disk radius.
So in the above equation, the dominant terms are those related to the density length scale.
We can see that 
\begin{equation}
\Omega_{a,r}^2 \approx k_{\phi}^2 c_a^2 + \frac{c_a^2 c_s^2}{c_a^2+c_s^2}\left(\frac{d\ln\rho}{d r} \right)
\left(\frac{d\ln\rho}{d r} \right) > 0 \ .
\end{equation}
From this eqution we can see that, when the instability ocurrs at either the density decreasing 
inner edge or the density increasing outer edge, the contribution of $\Omega_{a,r}^2$ is
always positive, that is, it will decrease the instability growth rate and stabilize 
the RVI.

\begin{figure}
\begin{center}
\epsfig{file=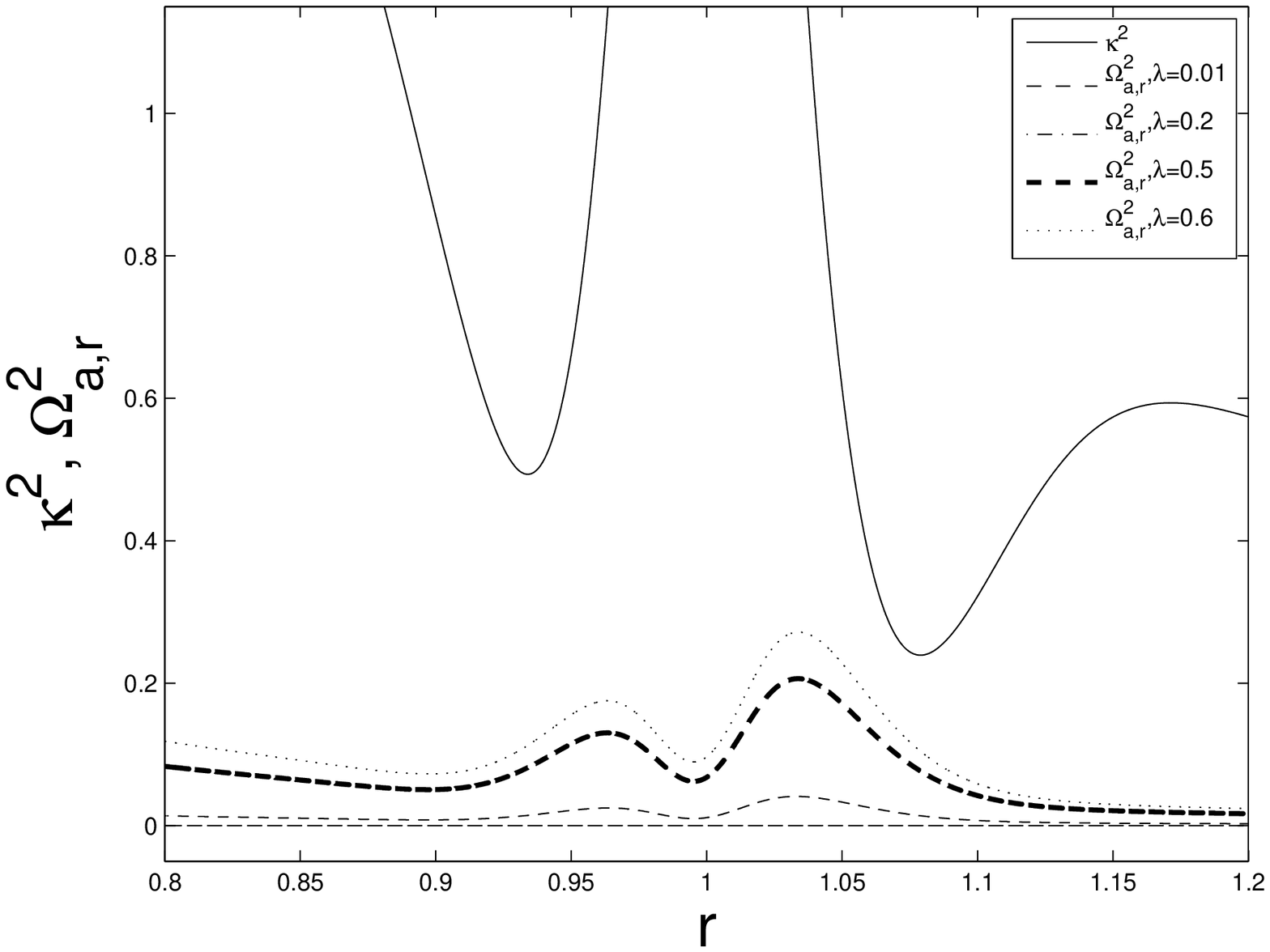,height=2in,width=5in,angle=0}
\end{center}
\caption{The radial profile of $\kappa^2$ and $\Omega_{a,r}^2$ of the $m=5$ outer edge mode
         for different magnetic field strength. As the magnetic field increases,
         $\Omega_{a,r}^2$ increases around where the excitation of instability
         takes place (both the inner edge and outer edge)
         and stabilizes the RVI.
        }
\label{kappa2omr2}
\end{figure}

In Figure \ref{kappa2omr2}, we give both $\kappa^2$ and $\Omega_{a,r}^2$ as a function of radius, 
we can see that stronger magnetic fields
give larger value of $\Omega_{a,r}^2$ which stabilizes the RVI.
To investigate the magnetic field gradient on the behavior of the instability, we also tried different
values of $\alpha = -1, 0, 1, 2$, 
the stabilizing effect of toroidal magnetic fields still holds.

\begin{figure}
\begin{center}
\epsfig{file=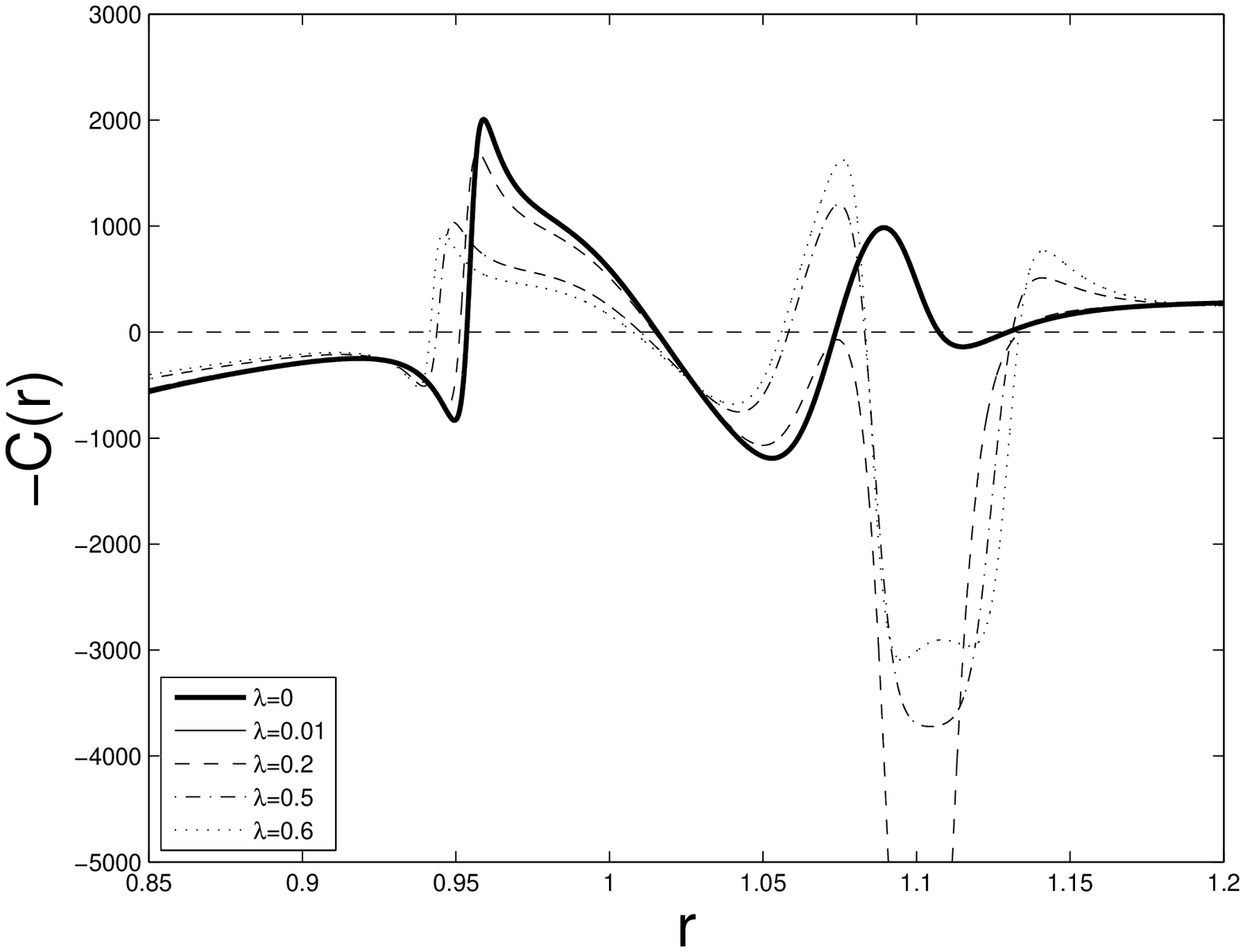,height=2in,width=5in,angle=0}
\end{center}
\caption{The ``potential well" around $r = 1.05$ becomes narrower and shallower
         with the increase of magnetic field strength. The excitation of the 
         RVI becomes weaker. 
         For small magnetic field $\lambda = 0.01$, it is indistinguishable from
         the nonmagnetic case.
        }
\label{lampotential}
\end{figure}

The basic physical mechanism for these modes is essentially the same as the hydrodynamic RVI 
(Lovelace 1999; Li et al. 2000).
Although in the nonaxisymmetric analysis, the ``potential'' $-C(r)$ becomes complex,
it is still of guidance to plot the real part of the ``potential'', $C(r)$ in equation (\ref{2ndorderode}). 
The real part of the function $-C(r)$ for different magnetic field strengths 
is shown in Figure \ref{lampotential}. 
For the outer edge mode, 
the negative ``potential well'' around $r/r_0=1.05$ is the unstable region.
When an unstable mode is excited in the ``potential well'' of the unstable region, the two positive
potential peaks at two sides of the unstable region cause this mode to be evanescent in such regions.
These two trapping regions would partially act as reflection boundary for the amplification to work,
although there will be a finite probability for the mode to tunnel through the ``potential barriers'' (as the 
``potential'' becomes again negative when it goes far from the unstable region).

The ``potential well" of $-C(r)$ in the wave equation can also 
confirm stabilizing effects of magnetic fields. 
Comparison of Figure \ref{kappa2omr2} and Figure \ref{lampotential} shows that
the increase of $\Omega_{a,r}^2$ at $r=1.05$ causes the  potential to
become narrower and shallower. 
This means that the excitation of the instability is weaker.
Interestingly, another potential well seems to have developed at 
$r = 1.1$, which may mean that there exists another branch of modes in our equations. But
we suspect that this branch of modes is stable and therefore is not the focus of this paper.

\subsubsection{Instability Threshold}
\begin{figure}
\begin{center}
\epsfig{file=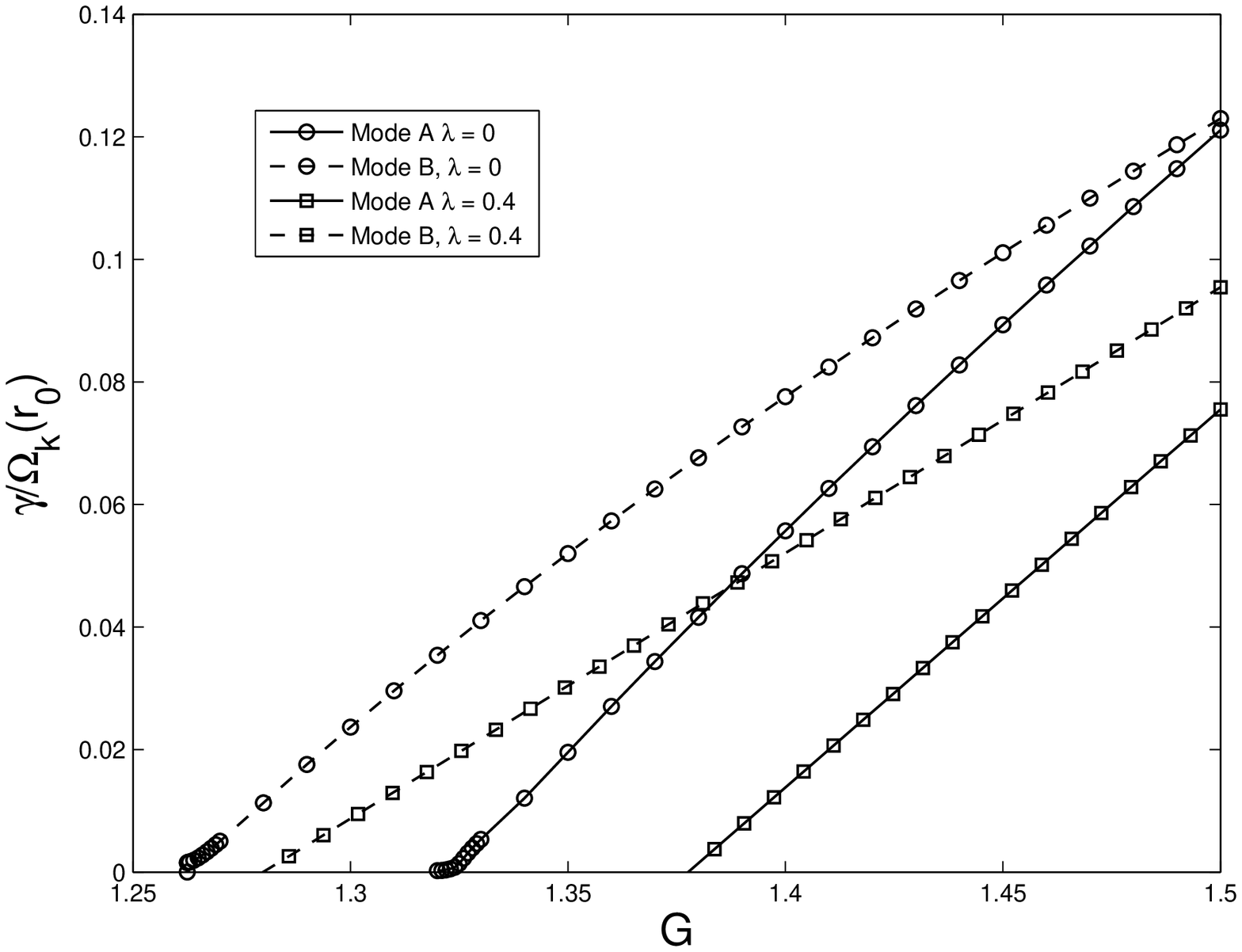,height=2in,width=5in,angle=0}
\end{center}
\caption{
Dependence of the mode growth rates on the amplitude of the surface density gap/dip 
$G$ for the $m=5$ unstable modes. The vanishing of the growth rate for 
$G<G_{\rm th}$ indicates the threshold for the RVI.
When magnetic fields are included, the threshold vlaues of $G$ increase. 
}
\label{threshold}
\end{figure}

Consider now the dependence of the growth rate on the dip/gap amplitude
$G$. As the amplitude $G$ decreases, the growth rate of the
instability is expected to decrease. 
Figure \ref{threshold} shows the calculation of the growth rate and mode frequency of hydrodynamic
Rossby vortex mode as a function of $G$ for
the $m=5$ unstable mode. The threshold value is $G_{\rm th} = 1.32$ and $1.26$ for the inner
edge and the outer edge mode, respectively. The threshold of the outer edge mode is
slightly lower than that of the inner edge mode. 
Note that the value of $G_{\rm th}$ depends on $\Delta $ and $c_s$, which are not discussed here.
For $\lambda = 0.4$, 
$G_{\rm th} = 1.38$ and $1.28$ for the inner edge and the outer edge mode, respectively.
When magnetic fields are included, $G_{\rm th}$ increases. 

\section{Conclusions}
\label{sec:diss}
We have carried out a 
linear analysis of the magnetized RVI associated with an axisymmetric, local radial density structure 
in a thin accretion disk. 
The flow is made unstable due to the existence of local extreme in the radial profile of the  
potential vorticity. 
Depending on the parameters, the
unstable modes are found to have substantial growth rates $\sim 0.1\Omega(r_0)$, 
where $r_0$ is the location of the surface density dip/gap.
We also find that strong toroidal magnetic fields would stabilize the RVI induced by the
radial density structures in disk. 
In situations where only weak magnetic fields exist, such as protoplanetry disks,
the existence of the magnetized RVI and their nonlinear outcome as vortices 
(e.g., Li et al. 2005)
indicate that the flows in the co-orbital region are more complicated than what is expected 
from the linear analysis. If the disk plasma $\beta$ is usually large (a few tens), the corresponding
$\lambda$ is about 0.3, from Figure \ref{lamformeq5}, we can see that the instability is only slightly
depressed by the toroidal magnetic fields and the RVI is still active even with the 
toroidal magnetic fields.

\acknowledgments
{\bf Acknowledgement:}
This research was supported by the Laboratory Directed Research and Development (LDRD) Programs
at Los Alamos and by the Institute for Geophysics and Planetary Physics (IGPP). C.Y. thanks the
support from National Natural Science Foundation of China (NSFC, 10703012).

\clearpage





\begin{thebibliography}{14}

\bibitem[Balbus \& Hawley]{bh91} Balbus, S. A. \& Hawley, J. F. 1991,
\apj., 376, 214

\bibitem[Balbus \& Hawley]{bh98} Balbus, S. A. \& Hawley, J. F. 1998, 
Rev. Mod. Phys., 70, 1




\bibitem[Chandrasekhar 1961]{Chandra61}
Chandrasekhar, S. 1961, Hydrodynamic and Hydromagnetic Stability (Oxford: Clarendon)

\bibitem[Curry \& Pudritz 1996]{curry96}
Curry, C. \& Pudritz, R. E. 1996, \mnras, 281, 119
                                          
\bibitem[de Val-Borro 2007]{devalborro07}
de Val-Borro, M., Artymowicz, P., D'Angelo, G., \& Peplinski, A. 2007,
\aap, 471, 1043

\bibitem[Gammie (1996)]{Gammie96}
{Gammie}, C. F. 1996, \apj, 457, 355

\bibitem[Goldreich et al. (1986)]{GGN86}
{Goldreich}, P., Goodman, J. \& Narayan, R. 1986, \mnras, 221, 339


\bibitem[Goldreich \& Tremaine (1980)]{GoldreichTremaine80}
{Goldreich}, P. \& {Tremaine}, S. 1980, \apj, 241, 425              










\bibitem[Li et al. 2000]{lietal00}
{Li}, H., Finn, J. M., Lovelace, R. V. E., \& Colgate, S. A. 2000, \apj,
533, 1023

\bibitem[Li et al. 2001]{lietal01}
{Li}, H., {Colgate}, S.~A., {Wendroff}, B., \& {Liska}, R. 2001, \apj,
551, 874

\bibitem[Li et al. 2005]{lietal05}
Li, H., Li, S., Koller, J., Wendroff, B. B., 
Liska, R., Orban, C. M., Liang, E. P. T.,\& 
Lin, D. N. C. 2005, \apj, 624, 1003





\bibitem[Lin \& Papaloizou 1986]{LinPap86}
Lin, D.~N.~C. \& Papaloizou, J. 1986, \apj, 309, 846


\bibitem[Lovelace et al. 1996]{Lovelace99}
Lovelace, R. V. E., Li, H., Colgate, S. A., \& Nelson, A. F. 1999, \apj, 513, 805










\bibitem[Ogilvie \& Pringle 1996]{ogilvie96}
Ogilvie, G. I., \& Pringle, J. E. 1996, \mnras, 279, 152

\bibitem[{{Papaloizou} \& {Pringle}(1984)}]{ppi84}
{Papaloizou}, J. \& Pringle, J. E. 1984, \mnras, 208, 721

\bibitem[Pino \& Mahajan]{pino09}
Pino, J. \&  Mahajan, S. M., astro-ph/0904.1633

\bibitem[Press et al. 1992]{press92}
Press, W. H., Teukolsky, S. A., Vetterling, W. T., \& Flannery, B. P. 1992,
Numerical Recipes, Cambridge Univ. Press, Cambridge



\bibitem[Tagger \& Pellat 1999]{tagger99} 
Tagger, M., \& Pellat, R. 1999, \aap, 349, 1003


\bibitem[Terquem 2008]{terquem08}
Terquem, Caroline E. J. M. L. J. 2008, \apj, 689, 532

%
                                
\bibitem[{{Ward}(1997)}]{Ward97}
{Ward}, W.~R. 1997, Icarus, 126, 261

\end{thebibliography}

\end{document}